\documentstyle[12pt,epsfig]{article}
\advance \topmargin by -\headheight
\advance \topmargin by -\headsep

\evensidemargin \oddsidemargin
\begin{document}
%               macros formatting and equations
\topmargin -.6in
\def\br{\begin{eqnarray}}
\def\er{\end{eqnarray}}
\def\be{\begin{equation}}
\def\ee{\end{equation}}
\def\({\left(}
\def\){\right)}
\def\a{\alpha}
\def\b{\beta}
\def\d{\delta}
\def\D{\Delta}
\def\g{\gamma}
\def\G{\Gamma}
\def\h{ {1\over 2}  }
\def\hp{ {+{1\over 2}}  }
\def\hm{ {-{1\over 2}}  }
\def\k{\kappa}
\def\l{\lambda}
\def\L{\Lambda}
\def\m{\mu}
\def\n{\nu}
\def\o{\over}
\def\O{\Omega}
\def\p{\phi}
\def\rh{\rho}
\def\s{\sigma}
\def\t{\tau}
\def\th{\theta}
\def\ii {\'\i  }

\begin{center} {\large {\bf  Supersymmetric variational energies
for the confined Coulomb system }}\footnotemark \footnotetext{PACS
No. 31.15.Q, 11.30.P -  Key words: Supersymmetric quantum
mechanics, confinement}
\end{center}
\normalsize
\vskip 1cm
\begin{center} {\it  Elso Drigo Filho}  \footnotemark
\footnotetext{elso@df.ibilce.unesp.br}  \footnotemark
\footnotetext{Work partially supported by CNPq} \\ Instituto de Bioci\^encias, Letras e Ci\^encias
Exatas, IBILCE-UNESP\\ Rua Cristov\~ao Colombo, 2265 -  15054-000 S\~ao Jos\'e do Rio Preto - SP\\
\vskip 1cm {\it Regina Maria  Ricotta } \footnotemark
\footnotetext{regina@fatecsp.br}\\ Faculdade de Tecnologia de S\~ao Paulo, FATEC/SP-CEETPS-UNESP
\\ Pra\c ca  Fernando Prestes, 30 -  01124-060 S\~ao Paulo-SP\\
Brazil\\
\vskip 1cm
\end{center} {\bf  Abstract}\\

The methodology based on the association of the Variational Method
with   \-Supersymmetric Quantum
Mechanics is used to evaluate the energy states of the confined hydrogen atom. \\

\noindent{\bf  Introduction}\\

Confined systems have motivated many studies in different areas of
physics. Nowadays it culminates with low dimensional systems of
technological applications like quantum dots in semiconductors,
\cite{Ren}. Several authors have worked in a variety of methods
aiming the energy eigenvalues of such systems,
\cite{Zhu}-\cite{Sinha2}. Particularly, the conventional
variational method has also been used,
\cite{Varshni1}-\cite{Varshni2} and references therein.

In this context the association of the  variational method applied
within the forma\-lism of Supersymmetric Quantum Mechanics, SQM,
seems to be another appropriate alternative  to tackle such
problem.

SQM has provided good results concerning different nonrelativistic
quantum mechanical systems, such as the exactly solvable,
\cite{Gedenshtein}-\cite{ Levai},   the partially solvable,
\cite{Drigo1}-\cite{ Drigo7}, the isospectral, \cite{Todorovic}
and the periodic potentials, \cite{Sukhatme}.  In particular, it
gave good results for the energy states of systems well fit by the
Hulth\'en, the Morse and the  screened Coulomb potentials,
\cite{Drigo3}-\cite{Drigo8},  which are all non-exactly solvable
potentials in 3-dimensions. The latter were studied using a
methodology based on the association of the variational method
with SQM. Its starting point is the association of an {\it Ansatz}
for the superpotential. Through the superalgebra  the wave
function is $\;\;\;$evaluated, the so-called trial wave function
containing the variational parameters, which will be varied until
the energy expectation value reaches its minimum.

In this letter this approach is applied to the 3-dimensional
confined hydrogen atom in order to get its  the energy states.
This system is analogous to the hydrogenic donor located at the
centre of a spherical $GaAs-(Ga,Al)As$ quantum dot. The results
obtained here for the $1s$, $2p$ and $3d$ states are very good
when compared to recent results obtained from other approximative
methods as well as numerical exact results,
\cite{Varshni2}.\\

\noindent {\bf The variational method associated to SQM}\\

Consider a system described by a given potential $V_1$. The associated Hamiltonian $H_1$ can be
factorized in terms of  bosonic operators, in $\hbar = c = 1$ units,
\cite{Sukumar}-
\cite{Cooper1}.

\be H_1 =  -{d^2 \o d r^2} + V_1(r) =  A_1^+A_1^-  + E_0^{(1)} \ee
where $ E_0^{(1)}$ is the lowest eigenvalue.  Notice that the
function $V_1(r)$ includes the barrier potential term. The bosonic
operators are  defined in terms of the so called superpotential
$W_1(r)$, \be A_1^{\pm} =  \left(\mp {d \o dr} + W_1(r) \right) .
\ee As a consequence of the factorization of the Hamiltonian
$H_1$,   the Riccati equation must be satisfied, \be
\label{Riccati} W_1^2 - W_1'=  V_1(r) - E_0^{(1)}. \ee Through the
superalgebra, the eigenfunction for the lowest state is related to
the superpotential $W_1$ by \be \label{eigenfunction} \Psi_0^{(1)}
(r) = N exp( -\int_0^r W_1(\bar r) d\bar r). \ee

It should be stressed that if the potential is non-exactly
solvable, the Hamiltonian is not exactly factorizable which means
that there is no  superpotential that satisfies the Riccati
equation. However, the Hamiltonian can be factorized in terms of a
superpotential giving rise to an effective potential that best
mimics the true potential. Thus, using the superalgebra we
evaluate the wave function which will depend on free parameters,
the variational parameters.

The variational method is an approximative technique to evaluate
the energy spectra of a Hamiltonian $H$ and, in particular, its
ground state. Its central point is the  search for an optimum
wave-function $\Psi(r)$ depending on a set of parameters,
$\{\mu\}$. This is called the trial  wave-function. The approach
consists in varying these parameters  in the expression for the
expectation value of the energy \be \label{energy} E =
{\int{\Psi_{\mu}^* H \Psi_{\mu} dr}\over {\int{\mid \Psi_{\mu}
\mid^2 dr}}} \ee until this expectation value reaches its minimum
value. This value is an upper limit of the energy level. Even
though this method is usually  applied to get the ground state
energy only, it can also be applied to get the energy of the
excited states.

Thus, the aim of the variational method is the acquisition of this
optimum wavefunction. Conventionally one proposes a variational
wavefunction depending on a set of parameters. At this crucial
point, however, our strategy is to use SQM to obtain this
function. Based on physical arguments, an {\it Ansatz} for the
superpotential is proposed and, through the superalgebra, the
trial wave function is evaluated, (equation
(\ref{eigenfunction})). By minimizing the energy expectation value
with respect to the free parameters introduced by the {\it Ansatz}
the minimum energy is found.

We stress that, in fact, by making an {\it Ansatz} in the
superpotential corresponds to be dealing with an effective
potential $V_{eff}$ that satisfies the Riccati equation, i.e., \be
V_{eff}(r) = \bar W_1^2 - \bar W_1'+ E(\bar\mu) \ee where $ \bar
W_1 = W_1(\bar\mu)$ is the superpotential that satisfies
(\ref{Riccati}) for
$\mu = \bar\mu$, the parameter that minimises the energy of eq.(\ref{energy}).\\

\noindent {\bf The Confined Coulomb Potential} \\

The radial Hamiltonian equation for the Coulomb Potential, written
in atomic units,  is given by
\be \label{Coulomb} H = - {d^2 \o
dr^2}  + {l(l+1) \o r^2}- {2\o  r}.
\ee
We use the variational
method associated to SQM in order to get the energy states of the
confined Coulomb Potential. As the Coulomb potential is symmetric,
the confinement is introduced by an infinite potential barrier at
radius $r = R$. Thus we make the following  {\it Ansatz} for the
superpotential \be \label{W} W(r) =  - {\mu_1 \o r} + {\mu_2 \o R
- r} + \mu_3 \ee which depends on $R$, the radius of confinement,
and three variational parameters, $\mu_1$, $\mu_2$ and $\mu_3$.
The first and the last terms are already known from the
non-confined case, \cite{Cooper1}. The second term deals with the
confinement, as shown below in the effective potential.

Substituting this superpotential in the associated Riccati equation we arrive at a confining
effective potential $V_{eff}$, i.e., a potential which is infinite at $r = R$.  It is given by
\br
\label{Veff}
V_{eff}(r) &=& \bar W^2 - \bar W' + E(\bar\mu) \\
  &=& {\mu_1(\mu_1-1) \o r^2} + {\mu_2(\mu_2-1)\o (R-r)^2} - {2\mu_1\mu_3 \o r} - {2\mu_1\mu_2 \o
r(R-r)} + {2\mu_2\mu_3 \o R-r} + {\mu_3}^2 + E(\bar\mu) \nonumber
\er which is clearly infinite at $r = R$, as expected for a
confining system. Notice that the effective potential is evaluated
for  the values of the set of parameters $\{\bar\mu\}$ that minimise the energy.

As mentioned before, our trial wavefunction for the variational method is obtained from the
superalgebra through equation  (\ref{eigenfunction}), using the superpotential given by the {\it
Ansatz} made in equation (\ref{W}). It is given by
\be
\Psi(\mu_1, \mu_2, \mu_3,r) \propto  r^{\mu_1} \; (R - r)^{\mu_2} \;e^{-\mu_3 r}.
\ee
It depends of  three free parameters, $\mu_1$,  $\mu_2$  and  $\mu_3$ and vanishes
at $r = R$.

The energy  is obtained by minimisation of the energy expectation
value with respect to the three parameters.  The equation to be
minimised is given by \be \label{energymu} E(\mu_1, \mu_2, \mu_3)
= {\int_0^{R} \Psi(\mu_1, \mu_2, \mu_3,r)  [- {d^2 \o dr^2} - {2\o
r}  + {l(l+1)\o r^2}] \Psi(\mu_1, \mu_2, \mu_3,r) dr \o \int_0^{R}
\Psi(\mu_1, \mu_2, \mu_3,r) dr}. \ee

After the minimization of (\ref{energymu}) with respect to the
parameters $\mu_1$, $\mu_2$ and $\mu_3$ we get $E(\bar\mu)$ which,
from now on, will be referred as $E_{VSQM}$.

The tables $1$, $2$ and $3$ below show the results for different
values of $R$ and $l$ and the comparison with the exact numerical,
$E_{EXACT}$, and conventional variational results, $E_{V}$,
contained in \cite{Varshni2}. The comparison is made through the
percentage errors, \be \delta_{VSQM} = {|E_{EXACT}-E_{VSQM}|\over
E_{EXACT}}\% \ee and \be \delta_{V} = {|E_{EXACT}-E_{V}|\over
E_{EXACT}}\%. \ee
\newpage
{\bf Table 1.}  Energy eigenvalues (in Rydbergs) and percentage
errors for different values of $R$ for the $1s$ state, ($l=0$). Comparison is
made with results from \cite{Varshni2}.\\
\vskip .3cm
\begin{tabular}{|l|c|c|c|c|c|} \hline
\multicolumn{1}{|c} {R} & \multicolumn{1}{|c|} {$E_{EXACT}$} &
\multicolumn{1}{|c|} {$E_V$} & \multicolumn{1}{|c|} {$\delta_{V}$}
& \multicolumn{1}{|c} {$E_{VSQM}$} &
\multicolumn{1}{|c|} {$\delta_{VSQM}$} \\
\multicolumn{1}{|c} {} & \multicolumn{1}{|c} {Ref.
\cite{Varshni2}} & \multicolumn{1}{|c} {Ref. \cite{Varshni2}} &
\multicolumn{1}{|c} {} & \multicolumn{1}{|c|} {  } &
\multicolumn{1}{|c|} {}  \\ \hline 0.1 & 937.986  & 937.999  &
0.00 & 940.688 & 0.29 \\ \hline 0.2 & 222.140 & 222.143 & 0.00 &
222.757
& 0.28 \\  \hline 0.3 & 93.185  & 93.187 & 0.00 & 93.434 & 0.27 \\
\hline
 0.4 & 49.268 & 49.269 & 0.00 & 49.397 & 0.26   \\ \hline
0.5 & 29.496 & 29.497 & 0.00 & 29.571 & 0.25 \\ \hline
 0.6 & 19.055 & 19.056 & 0.00 & 19.100 & 0.23 \\ \hline
 0.7 & 12.940 & 12.941 & 0.00 & 12.968 & 0.22 \\ \hline
 0.8 & 9.0868 & 9.0874 & 0.01 & 9.1055 & 0.21 \\ \hline
 0.9 & 6.5244 & 6.5249 & 0.01 & 6.5370 & 0.19 \\ \hline
1.0 & 4.7480  & 4.7484 & 0.01 & 4.7565 & 0.18 \\ \hline
 1.2 & 2.5386 & 2.5388 & 0.01 & 2.5425 & 0.15 \\ \hline
 1.4 & 1.2942 & 1.2943 & 0.01 & 1.2958 & 0.13 \\ \hline
 1.6 & 0.54262 & 0.54263 & 0.00 & 0.54318 & 0.10 \\ \hline
 1.8 & 0.06511 & 0.06512 & 0.02 & 0.06522 & 0.17 \\ \hline
2.0 & -0.25000  & -0.24990 & 0.04 & -0.25000 & 0.00 \\ \hline
 2.2& -0.46407 & -0.46376 & 0.07 & -0.46391 & 0.03 \\ \hline
 2.4& -0.61280 & -0.61216 & 0.10 & -0.61227 & 0.09 \\ \hline
 2.6& -0.7196& -0.71682 & 0.16 & -0.71697 & 0.14 \\ \hline
 2.8& -0.79333 & -0.79152 & 0.23 & -0.79186 & 0.19 \\ \hline
3.0& -0.84793  & -0.84523 & 0.32 & -0.84706 & 0.10 \\ \hline
3.2& -0.88781 & -0.88396 & 0.43 & -0.88545 & 0.27 \\ \hline
3.4& -0.91710 & -0.91184 & 0.57 & -0.91440 & 0.30 \\ \hline
3.6& -0.93870 & -0.93174 & 0.74 & -0.93575 & 0.32 \\ \hline
3.8& -0.95469 & -0.94571 & 0.94 & -0.95156 & 0.32 \\ \hline
4.0 & -0.96653  & -0.95518 & 1.17 & -0.96509 & 0.15 \\ \hline
\end{tabular}\\
\newpage

{\bf Table 2.}  Energy eigenvalues (in Rydbergs) and percentage
errors for different values of $R$ for the $2p$ state, ($l=1$). Comparison is made with
results from \cite{Varshni2}.\\
\vskip .3cm
\begin{tabular}{|l|c|c|c|c|c|} \hline
\multicolumn{1}{|c} {R} & \multicolumn{1}{|c|} {$E_{EXACT}$} &
\multicolumn{1}{|c|} {$E_V$} & \multicolumn{1}{|c|} {$\delta_{V}$}
& \multicolumn{1}{|c} {$E_{VSQM}$} &
\multicolumn{1}{|c|} {$\delta_{VSQM}$} \\
\multicolumn{1}{|c} {} & \multicolumn{1}{|c} {Ref.
\cite{Varshni2}} & \multicolumn{1}{|c} {Ref. \cite{Varshni2}} &
\multicolumn{1}{|c} {} & \multicolumn{1}{|c|} {  } &
\multicolumn{1}{|c|} {}  \\ \hline 0.4 & 116.896  & 116.925 & 0.02
& 117.038 & 0.12 \\ \hline 0.5  & 73.318   & 73.336 &  0.02&
73.406  & 0.12   \\ \hline 0.6 &  49.874  & 49.887 & 0.03 &
49.934 &  0.12 \\ \hline 0.8 &   26.879 &  26.886 & 003 & 26.911
& 0.12  \\ \hline 1.0 & 16.446  & 16.451 &  0.03 & 16.466 & 0.12 \\
\hline 1.2 &  10.893  &10.897   & 0.03 &  10.906 &  0.12
\\ \hline 1.4  &  7.6138  & 7.6160  & 0.03&  7.6225 &  0.12 \\
\hline 1.6 &  5.5295  & 5.5311  & 0.03 & 5.5358 &  0.11 \\ \hline
1.8 &  4.1308  & 4.1321  & 0.03 &  4.1355 &  0.11 \\ \hline 2 &
3.1520  & 3.1530 & 0.03 & 3.1555 & 0.11 \\ \hline 2.2  &2.4438   &
2.4445  & 0.03 &  2.4465 &  0.11 \\ \hline 2.4 & 1.9173   & 1.9178
& 0.03 & 1.9193  & 0.11  \\ \hline 2.6 & 1.5170   &  1.5173 &0.03
& 1.5185  & 0.10  \\ \hline 2.8  &  1.2068  & 1.2070  & 0.02 &
1.2080  & 0.10  \\ \hline 3.0 & 0.96250  & 0.96269 & 0.03 & 0.96346
& 0.10 \\ \hline 3.5  &  0.54239  &  0.54245 & 0.01& 0.54239  &
0.10   \\ \hline 4.0 & 0.28705   & 0.28706  & 0.00 & 0.28732  &
0.09  \\ \hline 4.5  & 0.12373   &  0.12375 & 0.01 & 0.12385 &
0.10  \\ \hline 5.0  & 0.01519   &  0.01528 & 0.59 & 0.01523  &
0.29  \\\hline 5.5  &  -0.05910  & -0.05887  & 0.38 &  -0.05909 &
0.02\\ \hline 6.0 & -0.11111  & -0.11069 & 0.38  & -0.11111 & 0.00
\\ \hline 6.5  & -0.14818   & -0.14748  & 0.47 &  -0.14816 & 0.01
\\ \hline 7.0  & -0.17496   & -0.17392  & 0.59 &  -0.17490 & 0.03
\\ \hline 7.5  & -0.19451   & -0.19304  & 0.75 & -0.19440  & 0.05
\\ \hline 8.0& -0.20890  & -0.20691 & 0.95 & -0.20882 & 0.04 \\
\hline
\end{tabular}
\newpage
{\bf Table 3.}  Energy eigenvalues (in Rydbergs) and percentage
errors for different values of $R$ for the $3d$ state, ($l=2$). Comparison is made with results
from
\cite{Varshni2}.\\
\vskip .3cm
\begin{tabular}{|r|c|c|c|c|c|} \hline
\multicolumn{1}{|c} {R} & \multicolumn{1}{|c|} {$E_{EXACT}$} &
\multicolumn{1}{|c|} {$E_V$} & \multicolumn{1}{|c|} {$\delta_{V}$}
& \multicolumn{1}{|c} {$E_{VSQM}$} &
\multicolumn{1}{|c|} {$\delta_{VSQM}$} \\
\multicolumn{1}{|c} {} & \multicolumn{1}{|c} {Ref.
\cite{Varshni2}} & \multicolumn{1}{|c} {Ref. \cite{Varshni2}} &
\multicolumn{1}{|c} {} & \multicolumn{1}{|c|} {  } &
\multicolumn{1}{|c|} {}  \\ \hline  0.4 & 116.896  & 116.925 &
0.02 & 117.038 & 0.12 \\ \hline 0.5   & 126.320  & 126.396  & 0.06
& 126.417  & 0.08  \\ \hline 1.0 & 29.935  & 29.950 & 0.05 &
29.958 & 0.08 \\ \hline 1.5  & 12.570   & 12.576 & 0.05 & 12.579 &
0.07  \\ \hline 2.0  & 6.6550   &  6.6583 & 0.05 & 6.6601  & 0.08
\\ \hline 2.5  & 3.9920   & 3.9938 &  0.05& 3.9950 & 0.08  \\
\hline 3.0  &  2.5856  & 2.5867  & 0.04 &2.5876 &  0.08
\\ \hline 3.5  &  1.7618  & 1.7624  &  0.04&  1.7631 & 0.07 \\
\hline 4.0  &  1.2427  &  1.2431 & 0.03 & 1.2437 &  0.08
\\ \hline 4.5  &  0.89752  &  0.89776 &  0.03& 0.89820  & 0.08  \\
\hline 5.0 & 0.65823  & 0.65836 & 0.02 & 0.65873 & 0.08\\ \hline
5.5  &  0.48681  &  0.48686 &  0.01& 0.48717 & 0.07  \\ \hline 6.0
&  0.36068  &  0.36074 & 0.02 & 0.36095  & 0.08  \\ \hline 6.5  &
0.26583  & 0.26583  & 0.00 &  0.26603 & 0.08 \\ \hline 7.0  &
0.19318  & 0.19318  & 0.00 &0.19333  & 0.08   \\ \hline 7.5  &
0.13666   &  0.13668 &0.02  & 0.13677  & 0.08  \\ \hline 8.0 &
0.09212   & 0.09216  & 0.05 &  0.09220 & 0.08  \\ \hline 9.0  &
0.02801  &  0.02815 & 0.50& 0.02805  &  0.14 \\ \hline 10.0 &
-0.01419  & -0.01390 & 1.98 & -0.01417 & 0.13 \\ \hline 11.0 &
-0.04275  & -0.04229  & 1.09 &  -0.04275 & 0.01  \\ \hline 12.0 &
-0.06250   &  -0.06181 &  1.11& -0.06250  & 0.00  \\ \hline 13.0 &
-0.07637  &  -0.07540 & 1.27 & -0.07637  & 0.00  \\ \hline 14.0  &
-0.08623 &  -0.08479& 1.66 & -0.08622 & 0.01  \\ \hline 15.0 &
-0.09328  & -0.09160 & 1.81 & -0.09327 & 0.01\\ \hline
\end{tabular}\\
\vskip .3cm At this point we stress that for $R\rightarrow \infty$
the SQM results also agree with the exact  non-confined problem,
which corresponds to the removal of the infinite barrier. In this
case the energy is exact and corresponds to \be \label{energia}E =
- {1\over N^2}\;,\;\;\;\; N = n + l + 1 \ee

The table $4$ below shows the energy eigenvalues, $E_{VSQM}$, for
the $1s$, $2p$ and $3d$ states, ($n=0$ and $l = 0, 1, 2$) for increasing values of $R$.
Notice the convergency towards the energy of the exact non-confined case,
given by equation (\ref{energia}).\\
\newpage

{\bf Table 4.}  Energy eigenvalues, $E_{VSQM}$, (in Rydbergs) for
different values of $R$ for the $1s$, $2p$ and $3d$ states, ($n=0$ and $l = 0, 1, 2$).
It is also shown the exact result from (\ref{energia}).\\
\vskip .3cm
\begin{tabular}{|r|c|c|c|} \hline
\multicolumn{1}{|c} {} & \multicolumn{1}{|c|}{} &
\multicolumn{1}{|c|}{} & \multicolumn{1}{|c|}{}\\
\multicolumn{1}{|c} {R} & \multicolumn{1}{|c|}{l = 0} &
\multicolumn{1}{|c|}{l = 1} & \multicolumn{1}{|c|}{l = 2} \\
\multicolumn{1}{|c|} {} & \multicolumn{1}{|c|} {$E_{R \rightarrow
\infty = -1.00000}$} & \multicolumn{1}{|c|} {$E_{R\rightarrow
\infty = -0.25000}$} & \multicolumn{1}{|c|} {$E_{R\rightarrow
\infty = -0.11111}$} \\  \hline
10 & -0.99985 & -0.23754 & -0.01417\\
\hline 15 &
-0.99998 & -0.24941 & -0.09327\\  \hline 20 & -1.00000 & 0.24990 & -0.10788\\
\hline 30 & -1.00000 & -0.24999 & -0.11103 \\ \hline 50 & -1.00000
& -0.25000 & -0.11111 \\\hline
\end{tabular}\\

\vskip 1cm {\bf Comments and conclusions}\\

When dealing with the variational method associated with SQM to
get the energy states of a quantum dot  we have used a confining
effective potential depending on three variational parameters ($\mu_1$,
$\mu_2$ and  $\mu_3$). This was achieved through an {\it Ansatz} for the
superpotential. Using the superalgebra the trial wave function was
evaluated which depends on these three parameters.

Varying the trial wave function with respect to these parameters
we found good results for increasing values of the radius $R$, for which the energy is negative,
when compared to results obtained from other approximative
variational method and exact numerical results, \cite{Varshni2}.
From the point of view of the number of variational parameters
involved in the evaluation this improvement was indeed expected,
since in ref. \cite{Varshni2} there are two of such parameters
present in the calculation.  Nonetheless we stress that the great
advantage of our method is the achievement of the trial wave function
through the {\it Ansatz} for the superpotential, equation (\ref{W}).  This allowed us a previous comparative
analysis between the original potential, contained in equation (\ref{Coulomb}), and the effective
potential, equation (\ref{Veff}), which approaches the infinite at the neighbourhood
of the barrier. Following this line of reasoning, the wave function, evaluated through
the superalgebra, equation (\ref{eigenfunction}), vanishes at the border
because it finds a potential barrier that increases until
becoming impenetrable at $r = R$. These border effects
become more perceptive for smaller values of $R$. For increasing
values of $R$, smaller will be the border effects so that the
variational results get better. In other words, for large values
of $R$ the variational SQM {\it Ansatz} provides fast converging
results as displayed in Table 4. In this way we recover the
results for the non-confined system, when $R \rightarrow \infty $.

In conclusion, we remark that the results presented here suggest
that the association of  the superalgebra of SQM with the
variational method provides an appropriate approach to analyse confined systems.\\


\newpage
\begin{thebibliography}{99}
\bibitem{Ren} S. Y. Ren, Solid. State Comm. {\bf 102} (1997) 479
\bibitem{Zhu} J. L. Zhu, J. J. Xiong, B. L. Gu, Phys. Rev. {\bf B 41} (1990) 6001
\bibitem{Chuu} D. S. Chuu, C. M. Hsiao, W. N. Mei, Phys. Rev. {\bf B 46} (1992) 3898
\bibitem{Porras} N. Porras-Montenegro, S. T. Perez-Merchancano, Phys. Rev. {\bf B 46} (1992) 9780
\bibitem{Parades} H. Parades-Gutierrez, J. C. Cuero-Yepez, N. Porras-Montenegro, J. Appl. Phys.
{\bf 75} (1994) 373
\bibitem{Filippo} S. De Filippo, M. Salerno, V. Z. Enolskii, Phys. Lett. {\bf A76} (2000) 240
\bibitem{Sinha1} A. Sinha, Int. J. Quant. Chem. {\bf 79} (2000)
267
\bibitem{Sinha2} A. Sinha, R. Roychoudhury and Y. P. Varshni, Can.
J. Phys. {\bf 78} (2000) 141 and {\it ibid},  {\bf 79} (2001) 939
\bibitem{Varshni1} Y. P. Varshni, J. Phys. B: At. Mol. Opt. Phys. {\bf 30} (1997) L589
\bibitem{Varshni2} Y. P. Varshni, Phys. Lett. {\bf 252A} (1999) 248
\bibitem{Gedenshtein} L. Gedenshtein and I. V. Krive, So. Phys. Usp. {\bf 28} (1985) 645
\bibitem{Levai} G. L\'evai, Lect. Notes in Phys. {\bf 427} (1993) 427, Ed. H. V. von Gevamb,
Springer-Verlag
\bibitem{Drigo1} E. Drigo Filho, Mod. Phys. Rev. {\bf A9} (1994) 411
\bibitem{Drigo7} E. Drigo Filho and R. M. Ricotta, Physics of Atomic Nuclei {\bf 61} (1998) 1836
\bibitem{Todorovic} G. Todorovi\'c, V. Milanovi\'c, Z. Ikoni\'c and D.
Indjin, Phys. Lett {\bf A279} (2001) 268
\bibitem{Sukhatme} A. Khare and U. Sukhatme, J. Math. Phys. {\bf 40} (1999) 5473
\bibitem{Drigo3} E. Drigo Filho and R. M. Ricotta, Mod. Phys. Lett. {\bf A10} (1995) 1613
\bibitem{Drigo6} E. Drigo Filho and R. M. Ricotta, Phys. Lett. {\bf A269} (2000) 269
\bibitem{Drigo8} E. Drigo Filho and R. M. Ricotta, Mod. Phys. Lett. {\bf A15} (2000) 1253
\bibitem{Sukumar} C. V. Sukumar, J. Phys. A: Math. Gen. {\bf 18} (1985) L57
\bibitem{Sukumar2} C. V. Sukumar, J. Phys. A: Math. Gen. {\bf 18} (1985) 2917
\bibitem{Drigo2} E. Drigo Filho and R. M. Ricotta,  Mod. Phys. Lett. {\bf A4} (1989) 2283
\bibitem{Cooper1} F. Cooper, A. Khare and U. P. Sukhatme, Phys. Rep. {\bf 251} (1995) 267;

\end{thebibliography}
\end{document}